%% file: ms.tex
\newif\ifproceedings\proceedingstrue
\documentclass{IEEEtran}

\newfont{\mycrnotice}{ptmr8t at 7pt}
\newfont{\myconfname}{ptmri8t at 7pt}

\def\maintitle{Seele's New Anti-ASIC Consensus Algorithm with Emphasis on Matrix Computation}
\def\papertitle{\maintitle} 

\usepackage{subcaption}
\usepackage{times}
\usepackage{hyperref}
\usepackage{breakurl}
\makeatletter
\g@addto@macro{\UrlBreaks}{\UrlOrds}
\makeatother

\usepackage{array,multirow}
\usepackage{xspace}
\usepackage[usenames,dvipsnames]{xcolor}
\usepackage{balance}
\usepackage{paralist}
\usepackage{tabu}
\usepackage{cleveref}
\usepackage[numbers,sort&compress,square]{natbib}
\usepackage{adjustbox}
\usepackage{array}
\usepackage{tabularx}
\usepackage{pifont}
\usepackage{wasysym}
\usepackage{pdflscape}
\usepackage{adjustbox}
\usepackage{booktabs}
\usepackage{array}
\usepackage{varwidth}
\usepackage{multirow}
\usepackage{tikz}
\usepackage{threeparttable}
\usepackage{rotating}
\usepackage{afterpage}
\usepackage[labelfont=bf]{caption}

\usepackage{booktabs}
\newcolumntype{P}[1]{>{\raggedright\arraybackslash}p{#1}}

\definecolor{linkcol}{rgb}{0,0,1}
\definecolor{citecol}{rgb}{0,0.5,0}
\definecolor{urlcol}{rgb}{0.3,0,0}

\renewcommand{\and}{\hspace{5mm}}

\usepackage{enumitem}

\input{01_macros}

\begin{document}

\title{\papertitle}

\ifproceedings{
\author{

Luke Zeng$^*$,
Shawn Xin$^*$,
Avadesian Xu$^*$,
Thomas Pang$^*$,
Tim Yang$^*$,
Maolin Zheng$^*$,
\\
\vspace{2mm}
$^*$SeeleTech Corporation @San Francisco\\
}
}\fi

\maketitle

\input{03_introduction}
\input{04_theory}
\input{05_evaluation}

\input{06_discussion}
\input{07_summary}



{\footnotesize
\bibliographystyle{abbrv}
\bibliography{ms}
}


\end{document}

%% file: 01_macros.tex
\newcommand\comment[1]{}

\newcommand\pow{PoW\xspace}

\def\first{({\it i})\xspace}
\def\second{({\it ii})\xspace}
\def\third{({\it iii})\xspace}



%% file: 03_introduction.tex
\section{Introduction}

\label{intro}

Since 2008, blockchains have been gaining increasing attention for its revolutionary innovations which may make radical changes in many fields: payments and money transfers, voting~\cite{FollowMyVote}, intellectual property and digital rights management, sharing economy~\cite{Lazooz,Arcade,LemonWay}, social media~\cite{Akasha,Steem}, supply chain management (HyperLedger)\cite{altoros}, energy management, government and public records~\cite{publicRecords1,publicRecords2}, and so on~\cite{blockchain-applications}. The philosophy of blockchain is \emph{decentralization} and \emph{disintermediation}:  multiple untrusted or semi-trusted parties can directly and transparently interact with each other without the presence of a trusted intermediary. This property makes blockchain particularly appealing to financial institutions suffering from huge middleman costs in settlements and other back office operations.
 
So far, despite many breakthroughs and improvements, blockchain, compared to its traditional counterparts, still faces major hurdles before widespread adoption including but not limited to stability, performance and scalability. As per these properties, consensus protocols are the corner stone and are closely linked to them. Based on the same consensus protocol, all nodes will agree on the same criteria to pack, verify and mine a block. A consensus protocol is the vital safeguard to guarantee the blockchain's health and legality: only legal blocks (meeting the criteria of consensus protocol) can be added to the blockchain while illegal ones will be rejected. Two key properties that a consensus protocol should have: \first completeness, legal requests from correct clients are eventually processed, and \second consistency, if an honest node accepts (or rejects) a value then all other honest nodes make the same decision. Consensus is not a new topic: the distributed systems community has extensively studied it for decades, and developed robust and practical protocols that can tolerate faulty and malicious nodes~\cite{pbft,lamport1998part}. However, these protocols were designed for closed groups and blockchains have a higher requirement for zero fault tolerance.

The Proof-of-Work (\pow) consensus algorithm first implemented by Bitcoin's blockchain requires all miners to find the solution to a hash puzzle. Then the first miner to find the solution will claim the winnership and get the mining reward. Due to the probabilistic and one-way transformation process with a nonce to its hash, this kind of \pow consensus algorithm works well in keeping Bitcoin's decentralized network consistent and secure. However, there exists a big issue to this kind of \pow consensus algorithm: heavy load in hash arithmetic results in rewards dominated by machines with hashrate advantage like GPUs and ASICs ($\sim$50 TH/S). This deeply discourages a great population of users from joining the mining with regular personal computers ($\sim$1-100 MH/S)~\cite{hashrate}. More importantly, only with ``decentralized distributed reward'', a robust and secure decentralized blockchain's peer-to-peer network can be formed and thrive\cite{seelePaper1}.

In this paper, we will present a new \pow consensus algorithm used in Seele's main-net, MPoW (Matrix-Proof-of-Work). Compared to Bitcoin’s PoW consensus algorithm, MPoW requires miners to compute the determinants of submatrices from a matrix constructed with $n$ hashes other than brute-force-hashing using a hash function to find the target. This paper will evaluate this algorithm’s compatibility with difficulty adjustment. Then we will discuss its efficiency in countering machines with hashrate advantage, and its feasibility to personal computers. We believe more innovative consensus protocols can be developed based on this algorithm.

%% file: 04_theory.tex
\section{Theoretical Foundation}

\label{Random 0/1 matrix}


\subsection{SHA function: security and feasibility}

A blockchain consensus algorithm must be secure and hard to compute but easy to verify. A secure hash algorithm (SHA) function can be informally defined as a function that maps a message of arbitrary length to a fixed length ($m$-bit) hash value, is easy to compute but hard to invert, and in which a collision, finding two different messages with the same hash value, is computationally infeasible.

Specifically, a strong cryptographic hash function $h$ is usually expected to satisfy the following requirements\cite{sha256}:

\first \textbf{Collision resistance} : it must be computationally infeasible to find any two distinct messages $M$ and $M’$ such that $h(M)$ = $h(M’)$. The best collision resistance one can hope to achieve with an $m$-bit hash function is upper bounded by the $O(2^{m/2})$ complexity of a birthday attack\cite{birthdayAttack}.

\second \textbf{Preimage resistance} (one wayness) : given the $h(M)$ hash value of an unknown message $M$, it must be computationally infeasible to find any message $M’$ (equal or not to M) such that $h(M’)$ = $h(M)$. The best preimage resistance one can hope to achieve with an $m$-bit hash function is upper bounded by the $O(2^m)$ complexity of an ``exhaustive'' search.

\third \textbf{Second preimage resistance} (weak collision resistance) : given any $M$ known message and its $h(M)$ hash value, it must be computationally infeasible to find any $M’$ message distinct from $M$ such that $h(M’)$ = $h(M)$. The best preimage resistance one can hope to achieve with an $m$-bit hash function is upper bounded by the $O(2^m)$ complexity of an “exhaustive” search.


Requirement \first is by far the most important one in practice for the assessment of any candidate hash function such as the one considered in this paper, since:

• A collision resistant hash function is necessarily second preimage resistant, i.e. \first$\rightarrow$\third ;

• Although some theoretical counter examples of the implication \first$\rightarrow$\second are easy to construct, for most algorithms in practice, the existence of a computationally feasible preimage search attack would automatically result in a computationally feasible collision search attack. Specifically, assuming a preimage search attack exists, to prove that collision search attack also exists, one just have to draw a sufficient number of $M$ messages from a sufficiently large set of messages, and then apply the preimage search attack to $h(M)$ until eventually an $M'$ preimage distinct from M is be found.

Thus, in order to assess the security of a candidate cryptographic hash function, such as the one analyzed in this paper, it is nearly sufficient to restrict oneself to the investigation of the collision resistance properties of the considered function and of the collision resistance properties on the underlying compression function.

\subsection{Random $0/1$ matrix}

As we discussed previously, the secure hash algorithms (SHAs) are designed to be one-way functions. Additionally, SHAs exhibit the avalanche effect, where the modification of very few letters to be encrypted causes a significant change in the output. Hence, SHAs will provide random results. If we use specific number of random hashes to construct a matrix, the matrix will be a ``random" matrix. Herein, we discuss more details about random 0/1 matrices.

For simplicity, we start with results on $2\times2$ matrices with iid uniformly distributed elements. Let 
\begin{equation}
A=\left[ \begin{array}{ll}{a_{11}} & {a_{12}} \\ {a_{21}} & {a_{22}}\end{array}\right]
\end{equation}
be a $2\times2$ matrix with elements $a_{ij}$ ($i, j$ = 1,2) where $a_{ij}$ are iid random variables with density,

if  \(x \in[0,1]\),
\begin{equation}
f_{a}(x) = 1
\end{equation}

if \(x \in elsewhere\),
\begin{equation}
f_{a}(x) = 0
\end{equation}
Let $D$ = $det A$. Then the probability of density of $D$ is:

if  \(x \in [-1,0)\),
\begin{equation}
f_D(x) = (x+1)(2-\log (x+1))
\end{equation}
\begin{equation}
+x\left[-\log (-x)+\sum_{k=1}^{\infty} \frac{(-1)^{k+1}}{k}\right.
\end{equation}
\begin{equation}
\times\left(\sum_{i=1}^{k} \left( \begin{array}{l}{k} \\ {1}\end{array}\right) \frac{(x-1)^{k-i}\left((-x)^{i}-1\right)}{1}\right.
\end{equation}
\begin{equation}
+(x-1)^{k} \log (-x) ) ]
\end{equation}

if  \(x \in(0,1]\),
\begin{equation}
f_{a}(x) = f_{a}(-x)
\end{equation}

if \(x \in elsewhere\),
\begin{equation}
f_{a}(x) = 0
\end{equation}

Using the expression given as above, Williamson shows a graph of probability density of $D$ (the determinant of a $2\times2$ random matrix with independent element uniformly distributed on [0,1])\cite{WILLIAMSON1988167}. The density graph of the determinants peaks at 0 with a symmetry between positive and negative determinants. Moreover, as determinant value distances from 0, the density drops exponentially. Komlos has studied the singularity of random matrices and has shown that if \(\xi_{i, j}\) ($i,j$ = 1,2,...) are iid with a non-degenerate distribution\cite{Komlos}, then 
\begin{equation}
\label{prob_with_det0}
\lim _{\eta \rightarrow \infty}P \left( \left| \begin{array}{cccc}{a_{11}} & {a_{12}} & {\cdots} & {a_{1 n}} \\ {a_{21}} & {a_{22}} & {\cdots} & {a_{2 n}} \\ {\vdots} & {\vdots} & {\vdots} & {\vdots} \\ {a_{m 1}} & {a_{m 2}} & {\cdots} & {a_{m n}}\end{array}\right|=0\right) = 0
\end{equation}

%% file: 05_evaluation.tex
\section{Results}
\label{results}

\begin{figure}
  \centering
    \includegraphics[width=0.515\textwidth]{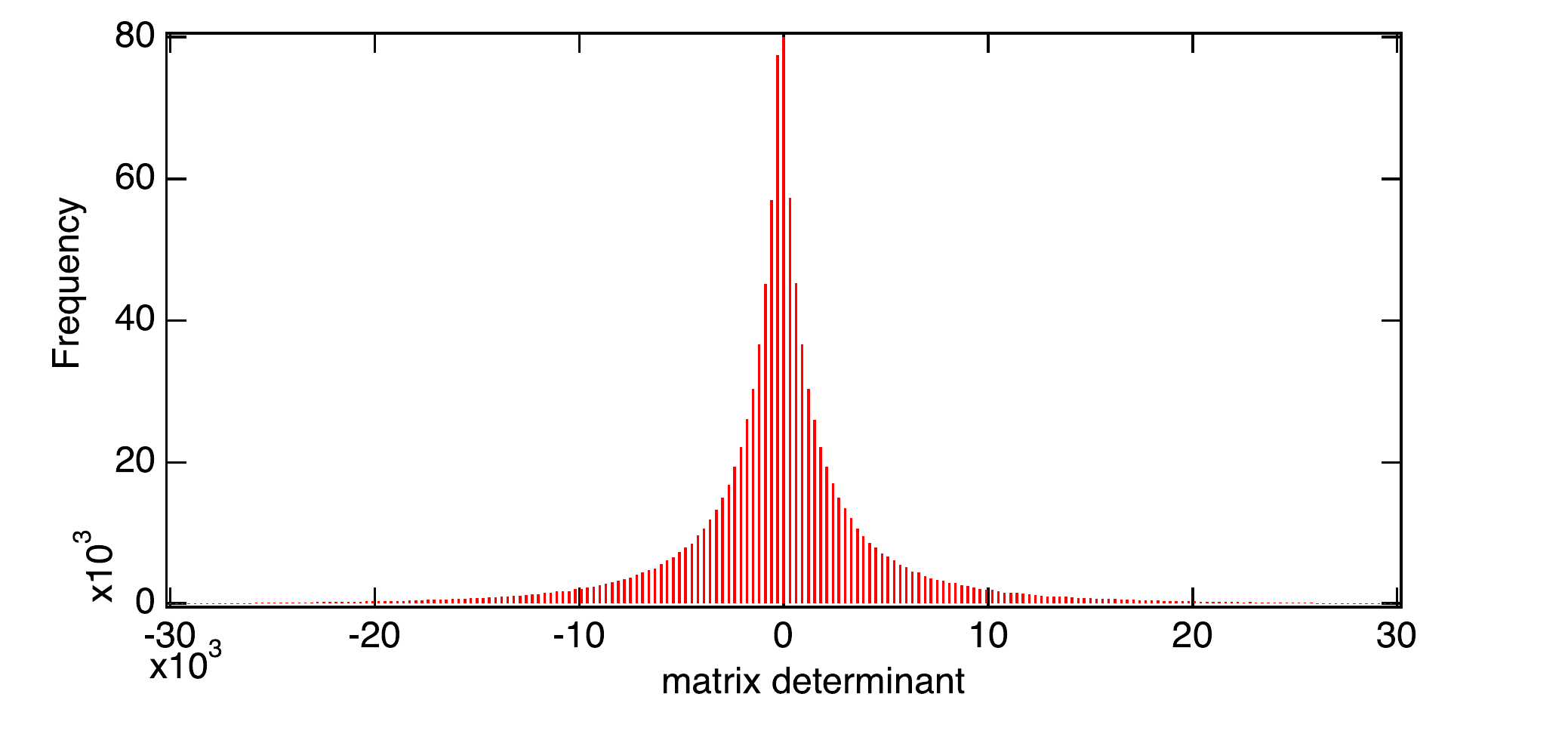}
    \caption{Determinant distribution of $30\times30$ (0/1)  matrix.}
    \label{det}
\end{figure}

In this section, we will show the results got from our Seele's main-net where we implement our new \pow consensus algorithm. Fig.\ref{det} presents the results on $30\times30$ matrices with iid uniformly distributed elements constructed with hashes. Similar to results in Williamson's paper\cite{WILLIAMSON1988167}, our histogram of determinants of $30\times30$ matrices (the size of data set is around 1 million) shows a peak at around 0, and as determinant value deviates from 0, the frequency decreases exponentially. All results agree well with the data from the forementioned paper.
For matrices constructed with specific number of hashes, we also observe that as the dimension increases, the frequency of matrix whose determinant is 0 decreases\cite{Komlos}. The singularity of a random matrix is out of scope of this paper, which we won't discuss in details here.

\begin{figure}
  \centering
    \includegraphics[width=0.515\textwidth]{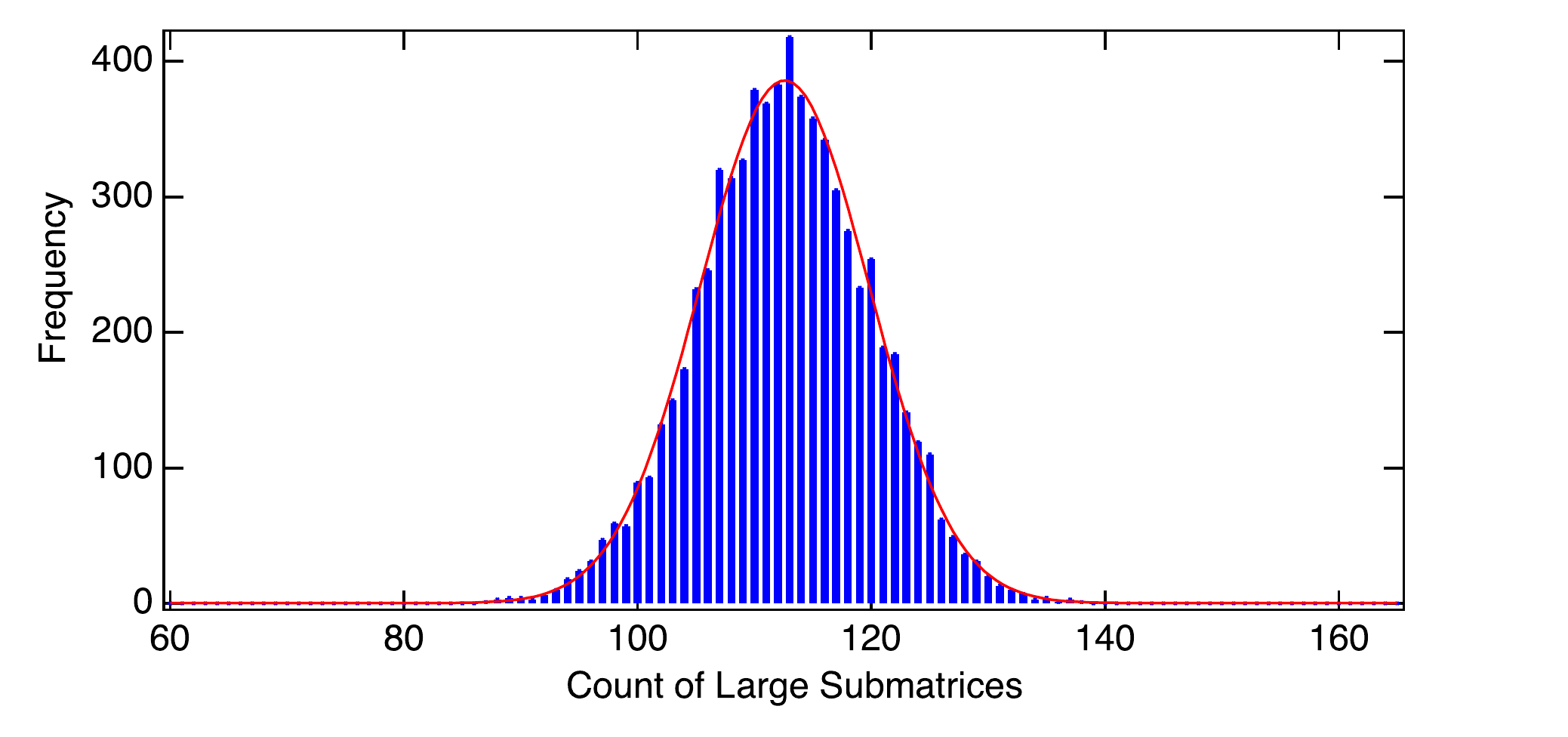}
    \caption{Distribution of number of ``large submatrices'' ($30\times30$ (0/1) matrix) subtracted from $30\times256$ (0/1)  matrices.}
    \label{non0count}
\end{figure}

As seen in Fig. \ref{det}, the distribution of determinant is symmetrical with a mean value around 0. In Fig.\ref{non0count},  we count the number of $30\times30$ submatrices with non-negative determinants, selected from $30\times 256$ matrices constructed by 30 hashes (for convenience, we call those submatrices as ``large submatrices''). Equation.\ref{prob_with_det0} shows that as the dimension goes up, the number of submatrices with 0 determinant decreases and approaches 0. When dimension equals 30, we can ignore the count of submatrices whose determinant is 0. Therefore, the probability of ``large submatrices'' will be close to 50\% and the number of ``large submatrices'' has the highest probability at 113 with a well defined normal distribution.

A (-1,1)-matrix having a maximal determinant is known as a Hadamard matrix \cite{Brenner and Cummings 1972}. The same bound of  $n^{n / 2}$ applies to such matrices, and sharper bounds are known when the size of the matrix is not a multiple of 4. A summary of what is known about such bounds is given by Orrick and Solomon\cite{matrixBound}.

For a (0,1)-matrix, Hadamard's bound can be improved to

\begin{equation}
|det A| \leq \frac{(n+1)^{(n+1) / 2}}{2^{n}}
\end{equation}

(Please refer to book \cite{Faddeev and Sominskii 1965}, problem 523; and paper\cite{Brenner and Cummings 1972}).

For an $n\times n$ (0,1)-matrix (i.e., a binary matrix), the largest possible determinants $\beta _n$ for $n$=1, 2, ... are 1, 1, 2, 3, 5, 9, 32, 56, 144, 320, 1458, 3645, 9477, ... (OEIS A003432\cite{OEIS A003432}). The numbers of distinct $n\times n$ binary matrices having the largest possible determinant are 1, 3, 3, 60, 3600, 529200, 75600, 195955200, 13716864000, ... (OEIS A051752\cite{OEIS A051752}).

%% file: 06_discussion.tex
\section{Discussion}


In this section, we will discuss the performance of our new MPoW consensus algorithm from the perspectives of block- time, difficulty adjustment and its efficiency in preventing machines with hashrate advantage, such as ASICs or GPUs, from dominating all mining rewards.
\subsection{Mining}
First of all, we will discuss the difficulty adjustment for our MPoW consensus algorithm. As more and more nodes (miners) join into the peer-to-peer network, stabilization of the network's mining process require a stale block time to guarantee safety and decentralization. On the other hand, the block difficulty adjustment of  MPoW algorithm needs to be gradual and smooth to achieve a stable block time (for Seele's main-net, the goal of block time is 10 $seconds$). As shown in the Fig.\ref{det} and Fig.\ref{non0count} respectively, there is a well-defined probability distribution for determinants of submatrices and for the number of ``large submatrices''. 

The difficulty adjustment formula is defined as follows:

\begin{equation}
\label{diff_equ}
d=d_{p}+\frac{d_{p}}{f} \times \max \left(1-\frac{t-t_{p}}{10},-99\right)
\end{equation}

Where: $d$ is the current block difficulty, $d_{p}$  is the previous block difficulty, $t$ is the current block time, $t_{p}$ is the previous block time, and $f$ is the smooth factor which is set to 2048 for Seele's main-net.

The block mining process will try to find a target matrix which meets two criteria:

\first The determinant $s$ of the submatrix constructed by first 30 columns and  30 rows meets:

\begin{equation}
s\geq d\times 65
\end{equation}

If the matrix meets the first criterion, in order to increase the calculation time percentage of mining a block, our MPoW consensus algorithm further requires the miner to calculate determinants of all submatrices and the number of the ``large submatrices'' to be as large as $count$.

\second $count$ is defined as:

\begin{equation}
count \geq \frac{256-n}{2} + \frac{n}{5}
\end{equation}
where $n$ is the row size of target matrix, here we use $n$ = 30.

\begin{figure}
  \centering
    \includegraphics[width=0.515\textwidth]{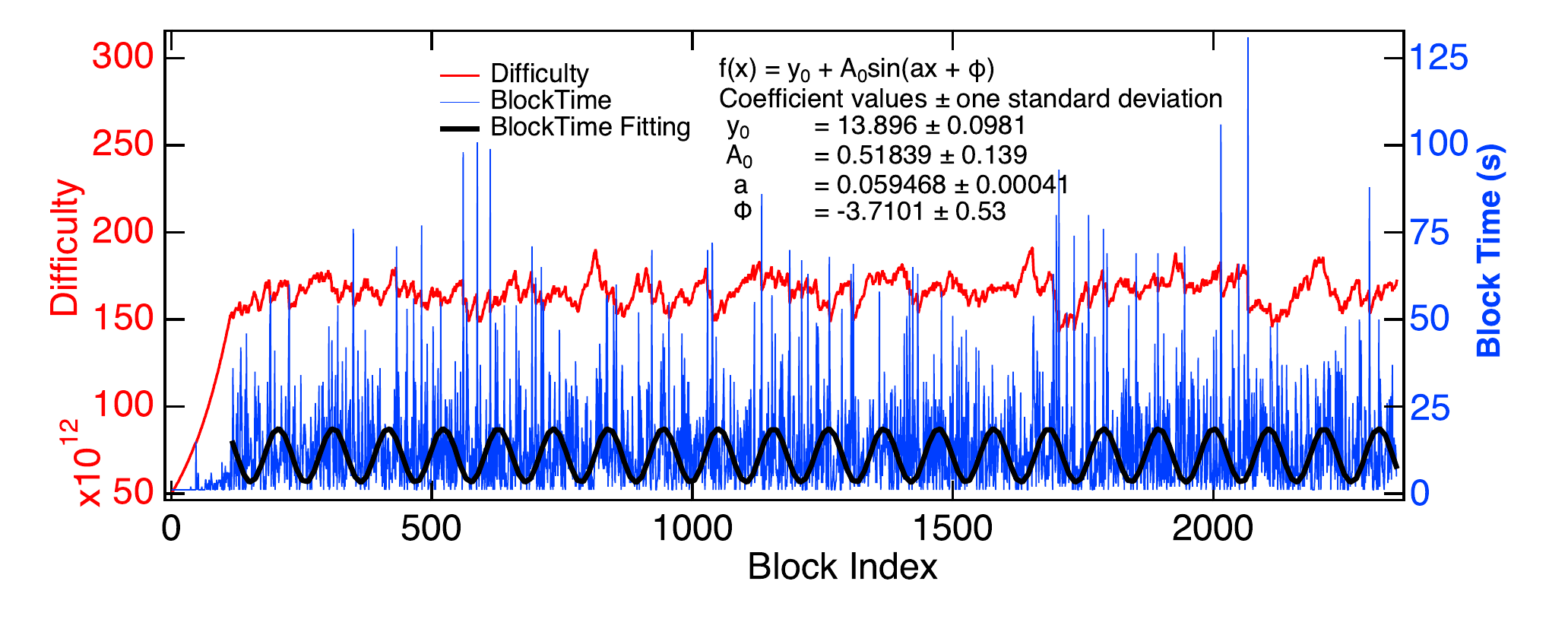}
    \caption{Block time distribution with dynamically adjusted difficulty. The black curve is the fitting result with a sinusoidal function.}
    \label{diff_time}
\end{figure}

In Fig.\ref{diff_time}, we present the difficulty and block time curves over block index. In order to better see the correlation between difficulty and block time, we intentionally set the smooth factor $f$ in Equation(\ref{diff_equ}) to 1. Generally, there is a strong correlation between the difficulty  (red curve) and block time (blue curve). The difficulty is subsequently adjusted to be smaller if the block time increases, vice versa. This indicates our difficulty function works quite well in dynamically adjusting the difficulty.  The target matrix finding process is feasible and within a reasonable block time. For a better look, we also fit the block time with a sinusoidal function as in Fig.\ref{diff_time}. The coefficient values from the fitting curve roughly give us an average block time of 14 seconds which well matches our 10 seconds block time goal considering other processes, such as packing a block, takes some extra time. Finally, note that if we set smooth factor $f$ to 2048 other than 1 (as we use in our Seele's main-net), the curves of difficulty and block time will be smoother with smaller deviations.

\subsection{Time Distribution}
\begin{figure}
  \centering
    \includegraphics[width=0.515\textwidth]{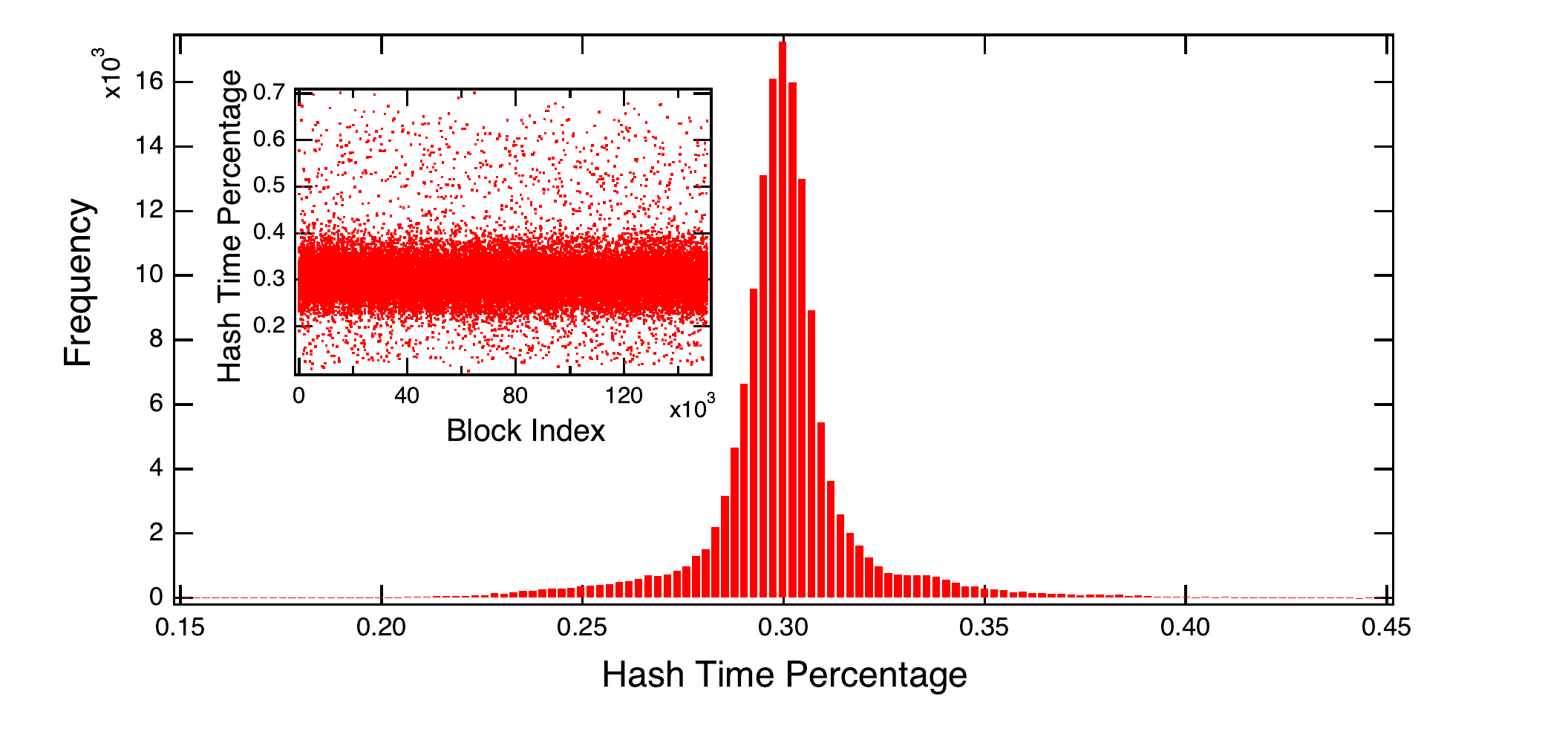}
    \caption{Hash time percentage during a Seele's block mining. The inset graph is the hash time percentage data.}
    \label{hashtime}
\end{figure}

Second, in the Fig.\ref{hashtime}, we show the results of hash time percentage within mining a block. The hash time percentage in the inset diagram of Fig.\ref{hashtime} shows a clear regression to 30\% with some random fluctuations. The histogram further confirms that the hash time percentage during mining a block is around 30\%. By comparison, traditional PoW consensus algorithms, requiring miner to solve a hash puzzle,  will take up almost 100\% of hash time when mining a block. However, hash time percentage is balanced out by our new MPoW consensus algorithm’s requirement: miner should further construct a matrix and calculate the determinants of its submatrices instead of just hashing. In this case, we can not only keep the blockchain safe and feasible with one-way data conversion using SHA function but also eliminate the advantage of machines with high hashrates.

%% file: 07_summary.tex
\section{Summary}
In this paper, we introduce a new PoW consensus algorithm (MPoW) which requires miner to use SHA function to get specific number, $n$ hashes and then use these hashes to construct a matrix($n\times256$) that satisfies two criteria: the determinant of first $n\times n$ submatrix should be not less than a target which can be dynamically adjusted based on the block time; Also, the number of submatrices with non-negative determinants should be larger than another given value. This MPoW consensus algorithm can efficiently eliminate the dominant advantage of machines whose hashrates are hundreds or thousand of times larger than personal computers. This consensus algorithm may pave the way to a real decentralizated blockchain. Furthermore, our new \pow (MPoW) consensus algorithm provides a new way to utilize the properties of a matrix to implement an efficient and secure blockchain's consensus algorithm. SeeleTech's research and development team will keep making an effort in this field and contributing to our community.\\